\documentclass[fleqn,10pt]{wlscirep}
\usepackage[utf8]{inputenc}
\usepackage[T1]{fontenc}

\usepackage{xcolor}
\usepackage{stmaryrd}
\usepackage{wasysym}
\usepackage{pifont}

\title{Editorial Trajectories in Wikipedia Reflect Underlying Hyperlink Structure}

\author[1]{Yeonji Seo}
\author[2, 3, *]{Mi Jin Lee}
\author[1, 2, *]{Seung-Woo Son}
\author[4, *]{Hang-Hyun Jo}
\author[5, 6, 7, *]{Yohsuke Murase}
\affil[1]{Department of Applied Artificial Intelligence, Hanyang University, 15588, Ansan, Republic of Korea}
\affil[2]{Department of Applied Physics, Hanyang University, Ansan 15588, Republic of Korea}
\affil[3]{Department of Physics, Pusan National University, Busan 46241, Republic of Korea}
\affil[4]{Department of Physics, The Catholic University of Korea, Bucheon 14662, Republic of Korea}
\affil[5]{RIKEN Center for Interdisciplinary Theoretical and Mathematical Science (iTHEMS), Wako 351-0198, Japan}
\affil[6]{RIKEN Center for Computational Science, Kobe 650-0047, Japan}
\affil[7]{Graduate School of Science and Engineering, Saitama University, Saitama 338-8570, Japan}
\affil[*]{mijinlee@pusan.ac.kr; sonswoo@hanyang.ac.kr; h2jo@catholic.ac.kr; yohsuke.murase@riken.jp}


\begin{abstract}
Wikipedia hyperlinks have primarily been studied as navigational tools for readers, but their role in how information providers move between articles during editing remains less explored. Here, we combine the hyperlink network among English Wikipedia articles with editorial histories to examine how article-to-article structure is associated with editors' transitions between articles. We first address the temporal aspect of edit transitions by showing that transitions between hyperlinked article pairs have shorter inter-event times (IETs) than those between non-hyperlinked pairs, indicating that connected articles are effectively closer in editing sequences. We then turn to the structural organization of editing behavior by coarse-graining the hyperlink network into 19 topical communities and measuring editors' topical diversity across them. Finally, we bring these temporal and structural views together by comparing each editor's transition network with the corresponding hyperlink subnetwork using Jaccard similarity. Combining these measures allows us to distinguish three editor types: \emph{`Specialists'} are characterized by focused editing within limited topical domains and transition patterns more closely aligned with the hyperlink structure (low topical diversity, shorter mean IETs, and higher Jaccard similarity), whereas \emph{`generalists'} cover broader topics and show weaker similarity to the hyperlink structure (high topical diversity, longer mean IETs, and lower Jaccard similarity). \emph{`Bots'} show a distinct algorithm-driven behavior, with low Jaccard similarity and the shortest mean IETs, a combination that departs from human-editor patterns despite their often high topical diversity. These findings demonstrate that the article-to-article hyperlink structure is not merely a static scaffold for reader navigation, but is observationally linked to the sequential organization of editorial activity in collaborative knowledge systems.
\end{abstract}
\begin{document}

\flushbottom
\maketitle

\thispagestyle{empty}

\section*{Introduction}
\label{sec:introduction}
Wikipedia is an online encyclopedia that aggregates a wide range of information, covers the same subjects across multiple languages, and operates as an open platform where anyone can edit~\cite{voss2005measuring}. Unlike conventional encyclopedias that are published upon completion, Wikipedia functions as a dynamic ecosystem characterized by continuous revisions~\cite{yasseri2012dynamics, ogushi2021ecology}. Its collaborative nature allows multiple editors to contribute to the same articles, thereby forming collective knowledge across nearly every field of knowledge~\cite{giles2005internet}. Within this collaborative framework, a key structural feature is the use of hyperlinks. Wikipedia forms a dense network of internal hyperlinks~\cite{son2012sampling, KUMMER201636, schwartz2021complex}, which connect related articles and provide navigational pathways for readers~\cite{lamprecht2017how}.

Previous studies have examined these hyperlinks primarily in the context of reader navigation. Such navigation has been studied in the ``Wikipedia game'', where players reach a target page using only hyperlinks~\cite{west2012human, helic2012analyzing, west2015mining, zhu2025milgram}, as well as through clickstream data~\cite{singer2017why, lamprecht2017how} to capture actual reader movement. Hyperlink structure itself has also been analyzed for information hierarchy~\cite{wei2014motif}, link prediction~\cite{kim2019anticipating, west2015mining}, and data mining~\cite{wei2015df}. Notably, research on the German Wikipedia has shown that higher centrality within the hyperlink network is positively correlated with both page length and the total number of editors~\cite{KUMMER201636}. These studies show that hyperlinks are not merely technical links between pages, but constitute a structural skeleton of knowledge~\cite{witten2008effective}.
In parallel, editing dynamics in Wikipedia have often been analyzed from the perspective of editor activity. Previous studies have characterized editing behavior in terms of editor activity levels~\cite{geiger2013using, yun2019early, ogushi2023comparison}, inter-event time (IET) patterns and bursty dynamics~\cite{kwon2016double, choi2021individual}, and editor-page bipartite networks that describe the overall Wikipedia ecosystem~\cite{ogushi2021ecology, shimada2023simple}. These approaches reveal who edited which page and how often, but they do not directly address how the page-to-page structure of Wikipedia is reflected in editors' transitions between articles. Thus, although hyperlinks have been extensively studied as navigational tools for readers, their relationship to the trajectories of information providers, namely editors, remains less understood.

Here, we ask whether the page-to-page hyperlink structure is associated with the temporal trajectories of Wikipedia editors. Editors are the ones who create and maintain hyperlinks, but the extent to which their own editing sequences overlap with this hyperlink structure has been little explored. In the vast information landscape of Wikipedia, hyperlinks may mark article pairs that are more readily considered together in editing activity, but quantitative evidence for this relationship remains limited. To address this issue, we combine the hyperlink structure with editorial history, rather than relying only on a bipartite representation of editors and articles.

\begin{figure}[b!]
\centering
\includegraphics[width=0.97\linewidth]{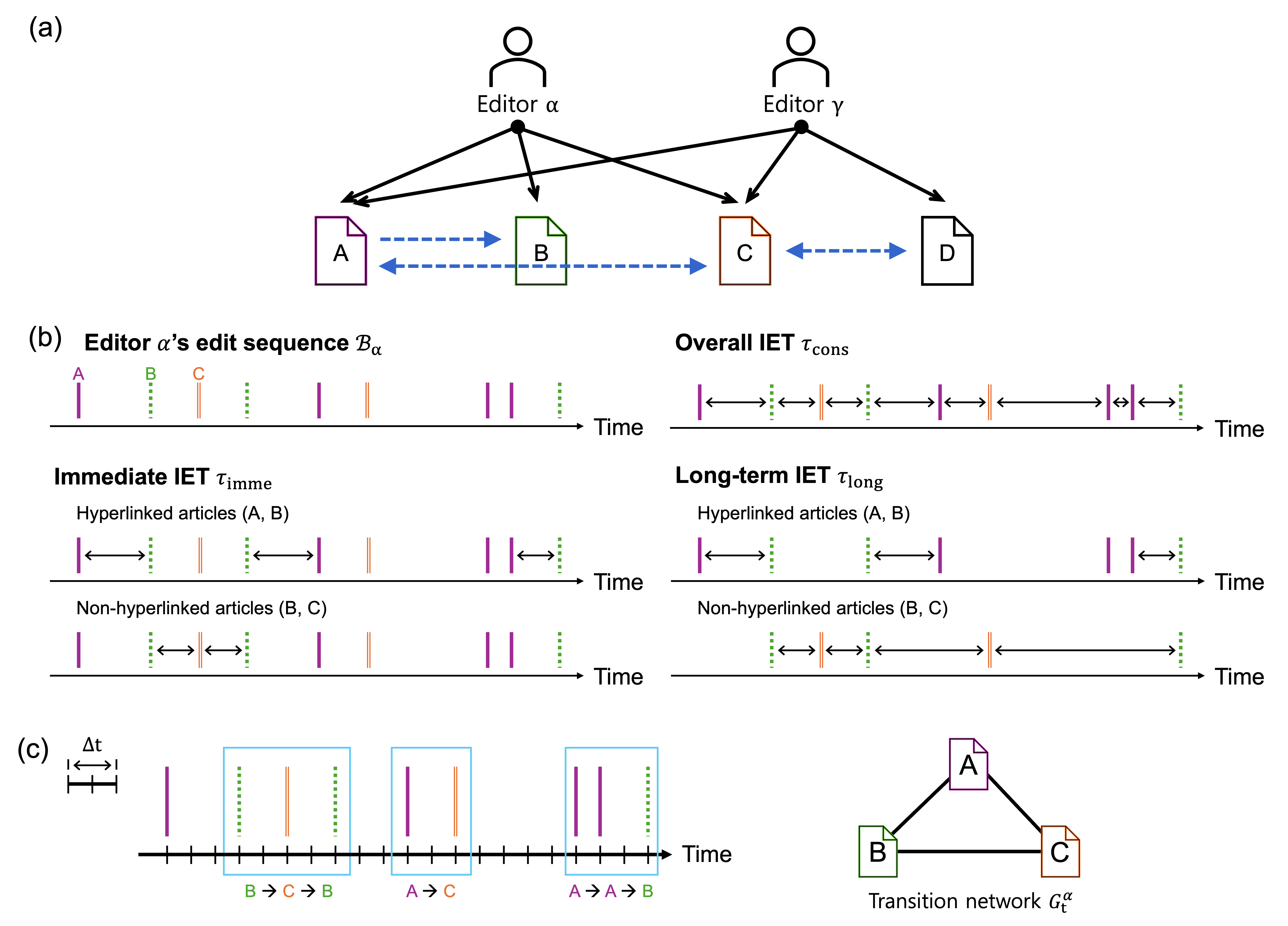}
\caption{
(a) Schematic representation of the editor-article relationship. Human and document icons correspond to editors and Wikipedia articles, respectively. Black solid arrows signify editing events, whereas blue dashed arrows connecting articles represent hyperlinks.
(b) Measurement of various types of 
IETs. The `edit sequence' of editor $\alpha$, $\mathcal{B}_\alpha$, is displayed at the top left, where vertical lines on the time axis denote edit timestamps for different articles. The measurement procedure for the overall IET $\tau_{\rm cons}$ is illustrated in the top right panel. The diagram further distinguishes between two specific types of transition: \emph{immediate} IET $\tau_{\rm imme}$ (bottom left) and \emph{long-term} IET $\tau_{\rm long}$ (bottom right). Within these categories, we compare cases involving hyperlinked articles (upper rows) and non-hyperlinked articles (lower rows). The IET being measured is explicitly represented by the left-right arrow ($\leftrightarrow$) in each scenario.
(c) Construction of the transition network $G_t^\alpha$. Consecutive edits separated by a time interval less than or equal to time window $\Delta t$ are grouped into a single edit sequence (indicated by solid sky-blue boxes). In the transition network $G_t^\alpha$, nodes represent articles and links represent edit transitions between articles.
}
\label{fig:Schematic}
\end{figure}

Using English Wikipedia's hyperlink data and edit history, we construct a unified structure that integrates editor-article connections with hyperlinks between articles, i.e., pages in the ``main'' \texttt{namespace} of the raw data, as shown in Fig.~\ref{fig:Schematic}. As illustrated in Fig.~\ref{fig:Schematic}(a), individual editors contribute to multiple articles, while each article is collaboratively edited by many editors. These articles are further connected by directed hyperlinks, and in some cases, two articles link to each other, forming reciprocal connections. For example, an editor $\alpha$ edits articles A, B, and C, some of which are also edited by another editor $\gamma$. While there is a directed hyperlink from A to B, articles A and C form a reciprocal hyperlink connection.

We examine the relationship between Wikipedia's hyperlink structure and editor trajectories from three connected perspectives. First, we quantify pairwise temporal proximity by comparing how often and how quickly editors move between hyperlinked and non-hyperlinked article pairs, using transition density and 
IETs. Second, we characterize the topical organization of editing by applying community detection to the hyperlink network and measuring whether editors concentrate on a narrow set of subject communities or distribute their edits broadly across them. Third, we compare each editor's transition network with the corresponding hyperlink subnetwork using the Jaccard similarity, thereby measuring the overlap between actual editing transitions and article-level hyperlink connections. This framework allows us to distinguish specialists, generalists, and bots in terms of topical diversity, temporal transition scale, and hyperlink-aligned editing behavior.

\section*{Results}
\label{sec:Result}
\subsection*{Difference of IET distributions between hyperlinked and non-hyperlinked article pairs}
\label{subsec:result_IET} 

\begin{table}[b!]
\caption{Statistics of IETs for different transition types. The `Count' and `Unique Pair' columns report, respectively, the total number of transition events and the number of distinct article pairs, both in millions ($10^6$), where the latter is computed from the union of all editors' unique pair sets. Reciprocally hyperlinked pairs are a subset of hyperlinked pairs. For non-hyperlinked pairs, the number of pairs is sampled for each editor to match that of hyperlinked pairs. Although pair numbers are matched at the individual editor level, the reported unique-pair counts are computed after taking the union over editors and therefore may differ between categories. Density is defined as the ratio of the total transition count to the number of unique pairs. Mean and median IET values are shown in appropriate time units (y: years, d: days, min: minutes).}
\label{tab:IET_stats}
\centering
\begin{tabular}{|l|r|r|r|r|r|}
\hline
\textbf{Category} & \textbf{Count ($10^6$)} & \textbf{Unique Pair ($10^6$)} & \textbf{Density} & \textbf{Mean IET} & \textbf{Median IET} \\ \hline
\emph{Consecutive} edits & 117.47 & - & - & 1.37 d & 7.28 min \\ 
\hline
\multicolumn{6}{|l|}{\emph{Immediate} transition between} \\ \hline
\quad hyperlinked articles & 16.94 & 3.37 & 5.03 & 1.13 d & 8.85 min \\ \hline
\quad reciprocally hyperlinked articles & 12.76 & 1.64 & 7.76 & 0.96 d & 6.77 min \\
\hline
\quad non-hyperlinked articles & 11.14 & 8.72 & 1.27 & 3.06 d & 91.93 min \\ \hline
\multicolumn{6}{|l|}{\emph{Long-term} transition between} \\ \hline
\quad hyperlinked articles & 1,918.82 & 11.98 & 160.16 & 1.06 y & 89.95 d \\ \hline
\quad reciprocally hyperlinked articles & 650.56 & 2.85 & 228.06 & 0.85 y & 52.93 d \\ \hline
\quad non-hyperlinked articles & 1,353.16 & 693.60 & 1.95 & 1.60 y & 200.33 d \\ \hline
\end{tabular}
\end{table}

\begin{figure}[ht]
\centering
\includegraphics[width=\linewidth]{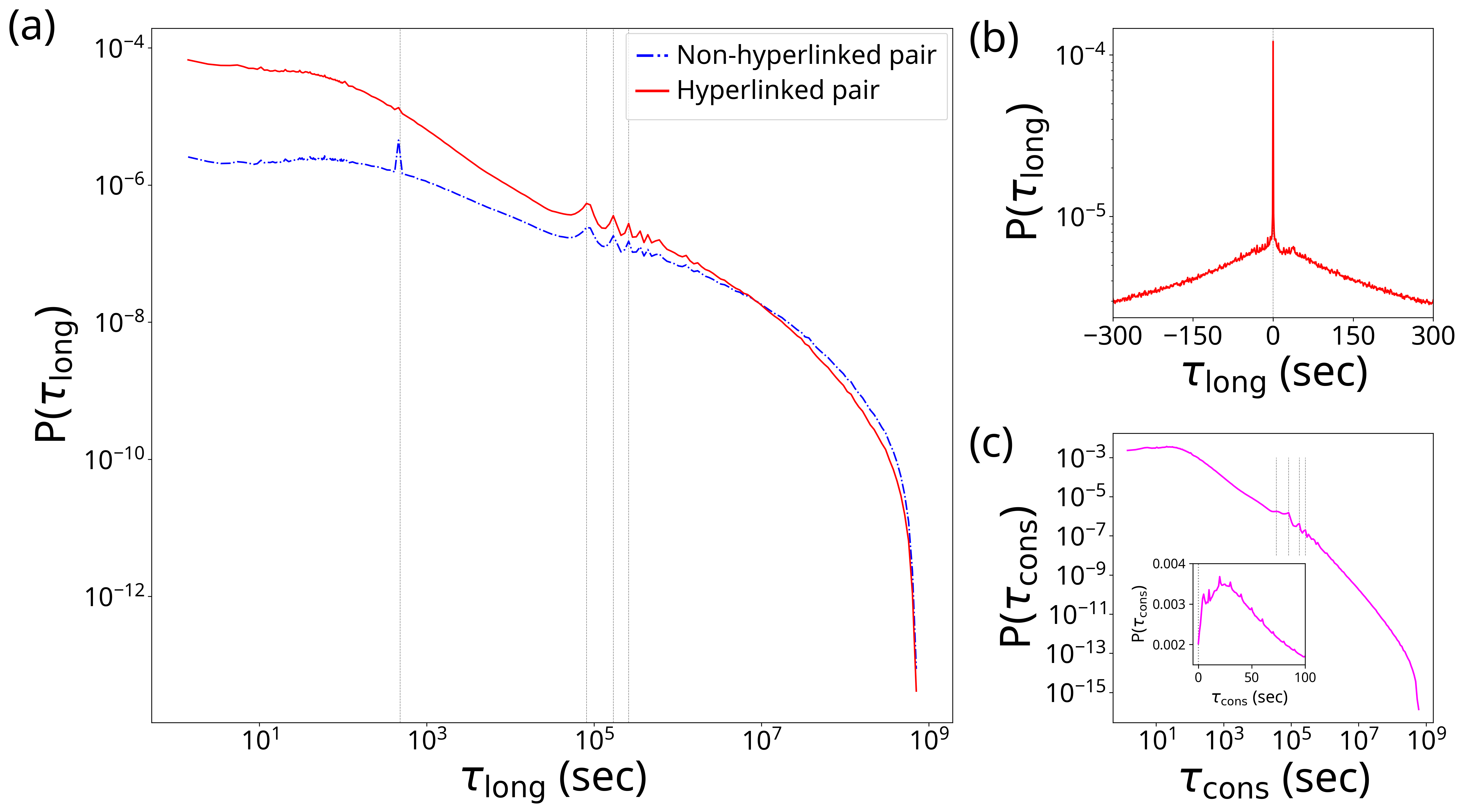}
\caption{Distributions of 
IETs for different transition types. (a) Distribution of long-term IETs, $P(\tau_{\rm long})$. The red solid and blue dashed lines represent hyperlinked and non-hyperlinked pairs, respectively. The gray vertical dotted lines indicate characteristic peaks at 484~s ($\approx 8$~min), $81\,100$~s ($\approx 1$~day), $170\,000$~s ($\approx 2$~days), and $259\,000$~s ($\approx 3$~days). The 484~s peak is a 2002 `Conversion script' artifact, unrelated to the circadian (daily) editing cycle. (b) Distribution of $\tau_{\rm long}$ only for uni-directional hyperlinks. Positive and negative values correspond to {`forward'} and {`backward'} transitions, respectively. The x-axis is restricted to $\pm 300$~s to highlight the symmetry near zero, and the gray vertical dashed line marks $\tau_{\rm long}=0$~s. (c) Distribution of overall IETs, $P(\tau_{\rm cons})$. The gray vertical dotted lines indicate peaks at $36\,600$~s ($\approx 10$~hours), $84\,400$~s ($\approx 1$~day), $175\,000$~s ($\approx 2$~days), and $266\,000$~s ($\approx 3$~days). The inset shows a zoomed-in view within 100~s on a linear scale, and the gray vertical dotted line marks $\tau_{\rm cons}=0$~s.
}
\label{fig:IET}
\end{figure}

To compare the temporal properties of editors' transitions between hyperlinked articles and those between non-hyperlinked articles, we examine how frequently such transitions occur and measure the associated 
IETs between edits in those pairs of articles. Here, we denote a two-article pair connected by a hyperlink as a `hyperlinked pair' and a pair without such a connection as a `non-hyperlinked pair' (the construction method is provided in the Methods section). For the analysis, we define three types of IETs [refer to Fig.~\ref{fig:Schematic}(b)]: (i) The IETs of \emph{consecutive} edits, termed overall IETs, $\tau_{\rm cons}$, are defined as time intervals between two consecutive edits by the same editor, irrespective of the articles edited. (ii) The IETs associated with \emph{immediate} transitions between two distinct articles, say $\beta$ and $\beta'$, are defined for each editor whenever the editor edits the article $\beta$ immediately after editing the article $\beta'$, i.e., without editing any other articles. We refer to these as immediate IETs, $\tau_{\rm imme}$. 
(iii) The IETs associated with \emph{long-term} transitions between two distinct articles, say $\beta$ and $\beta'$, are defined for each editor whenever the editor edits the article $\beta$ after editing the article $\beta'$, allowing the editor to edit any other articles. We refer to these as long-term IETs, $\tau_{\rm long}$. 
The detailed procedure for calculating IETs is provided in the Methods section.

The statistics of such IETs are summarized in Table~\ref{tab:IET_stats}. The total number of associated transition events is listed in the `Count' column, and the number of distinct article pairs appearing in the transitions is listed in the `Unique Pair' column. For a fair comparison at the individual editor level, we sample as many non-hyperlinked pairs as hyperlinked pairs for all articles edited by an editor. 
This controls for the imbalance in the number of possible non-hyperlinked pairs. As a result, the number of unique pairs is equal between the two categories at the individual editor level, while the number of transition events can differ. This sampling is applied to both immediate and long-term transitions. 

Table~\ref{tab:IET_stats} shows the statistics aggregated across all editors in the respective categories.
With equal numbers of pairs, transition events occur more frequently for hyperlinked pairs. Specifically, both immediate and long-term transitions between hyperlinked pairs occur approximately 1.5 times more frequently than those between the sampled non-hyperlinked pairs. For both transition types, the mean and median values of IETs are smaller for hyperlinked pairs than for non-hyperlinked pairs, indicating consistently faster transitions between connected articles. This effect is more pronounced for reciprocally linked pairs. These consistent patterns indicate that transitions between hyperlinked articles occur on shorter timescales than those between non-hyperlinked articles.

There is a significant gap between the mean and median values across all categories, signaling broad distributions of IETs. To illustrate these patterns, Fig.~\ref{fig:IET}(a) presents the distribution of long-term IETs, $\tau_{\rm long}$, as a representative example. The results indicate that editing transitions between hyperlinked pairs occur with significantly higher frequency in the small $\tau_{\rm long}$ regime, whereas a slight crossover is observed in the large $\tau_{\rm long}$ regime. These distributional characteristics are consistent with the observation that both the mean and median values of long-term IETs for hyperlinked pairs are substantially shorter than those for non-hyperlinked pairs in Table~\ref{tab:IET_stats}, suggesting that hyperlinked articles are effectively closer in the editing sequence.

To evaluate the influence of hyperlink directionality on editor behavior, we plot the probability distribution of $\tau_{\rm long}$ for uni-directional hyperlinks only, excluding reciprocal hyperlinks to isolate the direction effect, as shown in Fig.~\ref{fig:IET}(b). In this distribution, we introduce `negative' values of IET. The `positive' values correspond to `forward' transitions following the hyperlink direction, while the `negative' values represent `backward' transitions against it [see Eqs.~\eqref{eq:T_long,uni-link,forward}~and~\eqref{eq:T_long,uni-link,backward} in the Methods section]. The resulting distribution is nearly symmetric. This indicates that the statistical properties of IETs are largely independent of transition direction. A similar symmetric tendency is also observed in the immediate transition counts, although not shown here. This symmetry motivates the use of an \emph{undirected} transition network of articles [Fig.~\ref{fig:Schematic}(c)] in later analysis.

After the pair-based analysis, we examine the overall IETs, $\tau_{\rm cons}$, to characterize the temporal patterns of editors' activity independent of specific article pairs and to provide a baseline for comparison. Figure~\ref{fig:IET}(c) illustrates the distribution of $\tau_{\rm cons}$, measured by aggregating each editor's entire sequence of edit events across all articles. As shown in the inset of Fig.~\ref{fig:IET}(c), the occurrence of $\tau_{\rm cons}=0$ events is non-trivial, reflecting edits that occur within the temporal resolution of the data. Similar short-time events are also observed in Fig.~\ref{fig:IET}(b). These events can be attributed to factors such as simultaneous bot activity, rapid consecutive edits, or parallel editing of multiple articles. Here, bots are identified using the official labels provided in the raw data~\cite{wiki_bots} (see the Methods section).

A distinctive peak in the distribution of $\tau_{\rm long}$ is observed in Fig.~\ref{fig:IET}(a) at approximately 484~seconds (or 8~minutes). This peak is more prominent for non-hyperlinked pairs than for hyperlinked pairs, while it is not clearly observed in the $\tau_{\rm cons}$ distribution. This feature is attributed to a data artifact produced by a `Conversion script', a type of bot, during the February 2002 software migration. Specifically, it originates from the simultaneous resetting of timestamps to 15:43:11 and 15:51:15 on February 25, 2002. Although this migration event affects all articles globally, the resulting artifact becomes more pronounced for non-hyperlinked pairs due to their greater combinatorial diversity. This pattern becomes evident only when article pairs are separated by type, highlighting the importance of the pair-based analysis. We retain these events in the main analysis because bots are explicitly included as a distinct editor type in our framework. In addition, a weak peak is observed in the $\tau_{\rm cons}$ distribution at approximately 10~hours, which likely reflects typical resting periods of editors. Furthermore, circadian patterns with peaks at approximately 1, 2, and 3 days are observed in $\tau_{\rm cons}$~\cite{yun2016intellectual}, and similar patterns are present in $\tau_{\rm long}$.

\subsection*{Structural organization and community-based characterization of editors}
\label{subsec:result_Community}

\begin{figure}[b!]
\centering
\includegraphics[width=\linewidth]{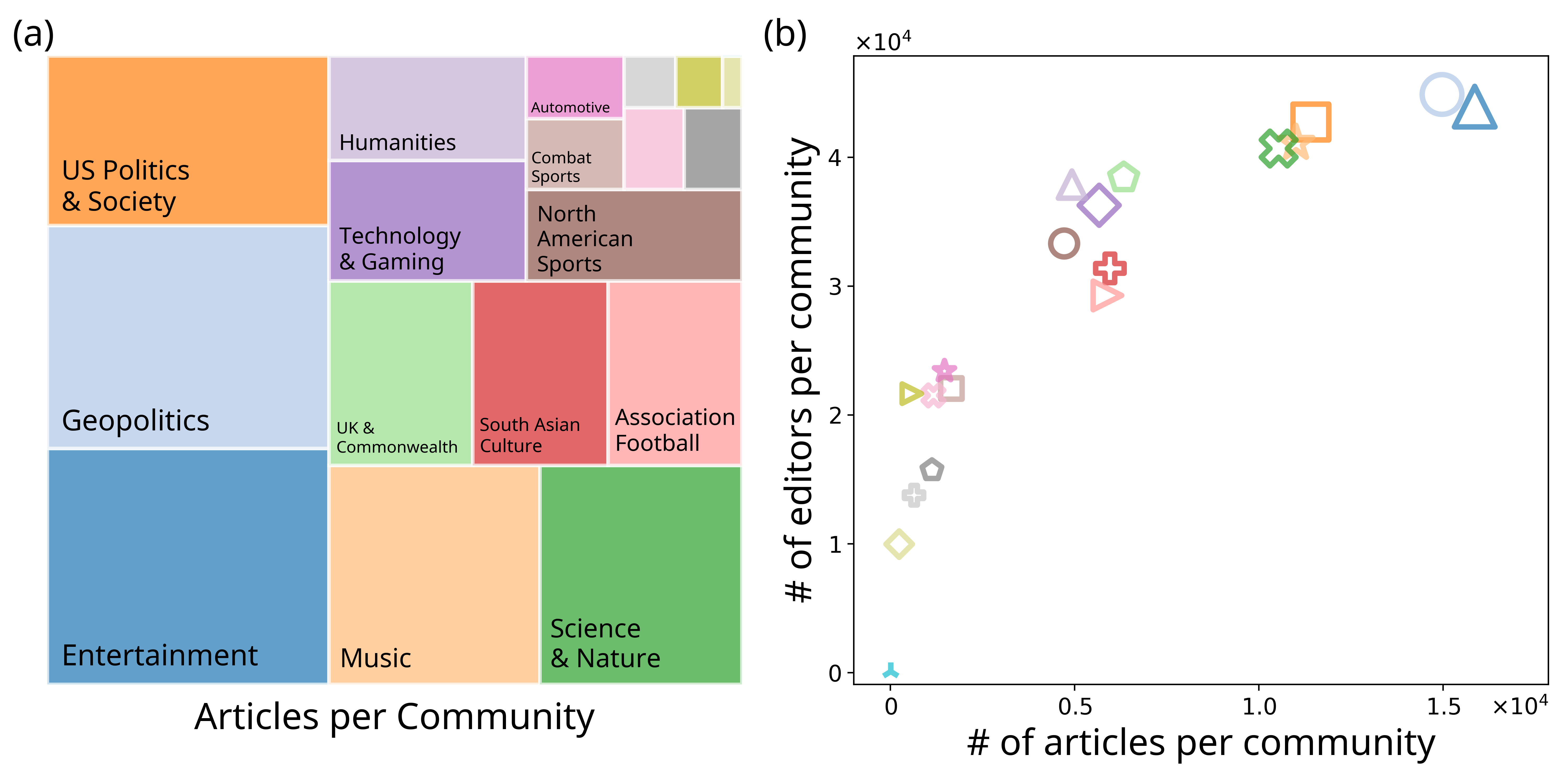}
\caption{ 
Visualization of the characteristics of the detected communities in the Wikipedia hyperlink network. (a) Treemap showing the distribution of articles (community size) across communities. (b) Scatter plot of community-level properties. The 19 distinct markers and colors correspond to the communities detailed in Table~\ref{tab:Community_information}. The x-axis represents the number $N_c^p$ of articles per community, and the y-axis represents the number $N_c^e$ of editors per community. The marker size is proportional to the total edit count $e_c$ within each community. All notations are defined in Table~\ref{tab:Community_information}.
}
\label{fig:Community}
\end{figure}

\begin{table}[ht]
\caption{
Overview of identified communities. This table summarizes the characteristics of 19 distinct communities found in the Wikipedia hyperlink network.
The `Colored Marker' column facilitates the identification of the data points shown in Fig.~\ref{fig:Community}. The same markers are used in Fig.~\ref{fig:Community}(b). 
The `Keyword' was extracted from the titles of the articles constituting each community with the assistance of a generative artificial intelligence, a large language model. 
Here, $e_c$ denotes the total number of edits in community $c$, $N_c^p$ denotes the number of articles constituting community $c$, and $N_c^e$ represents the number of unique editors who have edited articles in community $c$. Communities are sorted in descending order of $N_c^p$.
}
\label{tab:Community_information}
\centering
\begin{tabular}{|l|l|l|l|l|l|}
\hline
Index& Colored Marker& Keyword& $e_c$& $N_c^p$& $N_c^e$\\
\hline
1 & Blue $\triangle$ & Entertainment & $17\,276\,034$ & $15\,860$ & $43\,929$\\
\hline
2 & Light Blue $\bigcirc$ & Geopolitics & $18\,640\,258$ & $14\,972$ & $44\,872$\\
\hline
3 & Orange $\square$ & US Politics \& Society & $13\,397\,373$ & $11\,408$ & $42\,747$\\
\hline
4 & Light Orange \ding{73} & Music & $11\,655\,329$ & $11\,008$ & $41\,112$\\
\hline
5 & Green $\times$ & Science \& Nature & $11\,902\,121$ & $10\,537$ & $40\,677$\\
\hline
6 & Light Green $\pentagon$ & UK \& Commonwealth & $6\,714\,908$ & $6\,331$ & $38\,408$\\
\hline
7 & Red $+$  & South Asian Culture & $5\,533\,242$ & $5\,953$ & $31\,387$\\
\hline
8 & Light Red $\triangleright$ & Association Football & $7\,143\,851$ & $5\,890$ & $29\,276$\\
\hline
9 & Purple $\lozenge$ & Technology \& Gaming & $5\,666\,432$ & $5\,666$ & $36\,273$\\
\hline
10 & Light Purple $\triangle$ & Humanities & $6\,234\,959$ & $4\,926$ & $37\,879$\\
\hline
11 & Brown $\bigcirc$ & North American Sports & $5\,031\,423$ & $4\,715$ & $33\,309$\\
\hline
12 & Light Brown $\square$ & Combat Sports & $2\,071\,939$ & $1\,650$ & $22\,067$\\
\hline
13 & Pink \ding{73} & Automotive & $1\,523\,193$ & $1\,467$ & $23\,405$\\
\hline
14 & Light Pink $\times$ & Aviation & $1\,487\,253$ & $1\,168$ & $21\,518$\\
\hline
15 & Gray $\pentagon$ & Philippines \& Telenovelas & $985\,497$ & $1\,127$ & $15\,719$\\
\hline
16 & Light Gray $+$ & Tennis & $818\,793$ & $637$ & $13\,784$\\
\hline
17 & Olive $\triangleright$ & Chronology & $1\,072\,589$ & $580$ & $21\,655$\\
\hline
18 & Light Olive $\lozenge$ & COVID-19 Pandemic & $368\,118$ & $237$ & $9\,977$\\
\hline
19 & Cyan $\Yup$ & Meteorology & $1\,594$ & $2$ & $98$\\
\hline
\end{tabular}
\end{table}

While the IET analysis provides insights into the temporal aspects of editorial behavior based on transitions between article pairs, it does not directly capture the structural organization of articles within Wikipedia. To complement this temporal perspective, we next examine the structural organization of articles by constructing the hyperlink network.

We construct the Wikipedia hyperlink network, $G_h$, by defining articles as nodes and hyperlinks as directed links, which reveals how articles are interconnected to form a structured map of knowledge. To retain the fundamental connectivity between articles despite multiple transitions between the same pairs, we consider a directed and unweighted structure. The resulting network consists of 104K articles (nodes) and 14M hyperlinks (links), and its degree distributions are shown in Supplementary Fig.~S1. To reduce the complexity of article-level interactions and analyze editorial behavior at the topical level, we apply the Leiden algorithm~\cite{traag2019louvain} for community detection, yielding 19 communities with a modularity of 0.5806, indicating a well-partitioned structure. 

The information for each community is summarized in Table~\ref{tab:Community_information}. We extract representative keywords for the articles in each community with the assistance of a generative artificial intelligence, a large language model, 
confirming that the communities correspond to clearly distinguishable subjects such as `Entertainment', `Geopolitics', and `Science \& Nature'. Referring to Table~\ref{tab:Community_information} and Fig.~\ref{fig:Community}, we observe that communities with more articles tend to involve more editors and higher edit frequencies. The substantial variation in the numbers of articles, editors, and edits across communities reflects strong heterogeneity across subjects. This well-partitioned structure provides a foundation for analyzing editorial behavior at a coarse-grained level.

\begin{figure}[ht]
\centering
\includegraphics[width=\linewidth]{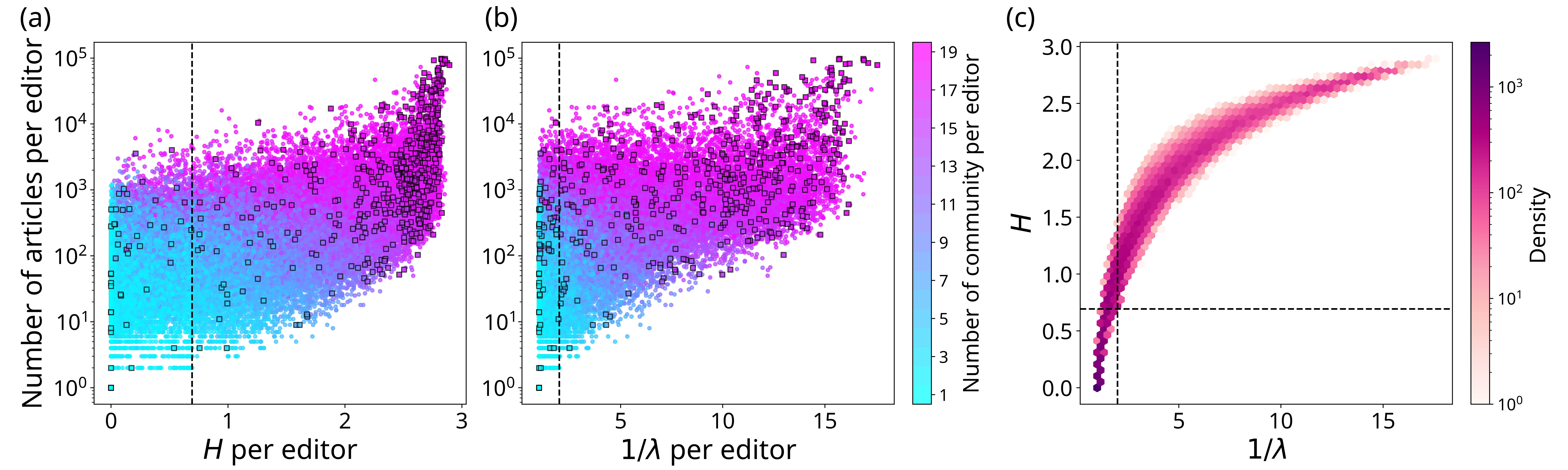}
\caption{Community-level diversity of editing patterns for individual editors. (a, b) Scatter plots of editor activity, defined as the number of distinct articles edited by an editor, versus diversity, measured by (a) entropy $H$ and (b) the inverse Simpson index $1/\lambda$. Each point corresponds to an editor 
and is colored according to the number of participating communities. Black-edged squares in (a)-(b) indicate editing bots. The dashed lines mark the classification baselines: (a) $H^*=\ln{2}$ and (b) $1/\lambda^*=2$. (c) Density plot of entropy and the inverse Simpson index. 
Color encodes the density of editors on a logarithmic scale. The vertical and horizontal dashed lines indicate $1/\lambda^*$ and $H^*$, respectively.
}
\label{fig:Entropy}
\end{figure}

Based on the derived community structure, we examine how editors utilize Wikipedia’s knowledge map to move between subjects at a coarse-grained level. To this end, for each editor $\alpha$, we define $G_h^{\alpha}$ as the induced subgraph of $G_h$ on the set of edited articles and map them to their corresponding communities, allowing us to characterize how concentrated or diverse their editing interests are. We quantitatively measure the diversity of editor interests using information entropy, $H_\alpha$, and the inverse Simpson index, $1/\lambda_\alpha$, defined for editor $\alpha$, as given in Eqs.~\eqref{eq:entropy}~and~\eqref{eq:inverse_Simpson_index} in the Method section. While both measures quantify diversity, they emphasize different aspects of the distribution: entropy reflects the overall heterogeneity of the distribution, including contributions from less frequent categories, whereas the inverse Simpson index is dominated by the most frequent elements and can be interpreted as the effective number of dominant communities. Specifically, a low entropy value indicates that an editor’s activity is unevenly distributed across communities, whereas a high value suggests a more uniform distribution. Meanwhile, the inverse Simpson index reflects the effective number of dominant communities: a lower value indicates that activity is concentrated in a small number of main communities, whereas a higher value corresponds to contributions spread across multiple dominant communities.

Figure~\ref{fig:Entropy} presents the calculated metrics for all editors. Defining the activity level as the number of articles edited by an editor, we observe that editors span a wide range of entropy and inverse Simpson index values at a given activity level. This indicates that even highly active editors can either concentrate on a small effective number of subject communities (low $H_\alpha$ and $1/\lambda_\alpha$) or cover a diverse range of subjects (high $H_\alpha$ and $1/\lambda_\alpha$) [see Supplementary Fig.~S2 for detailed density plots]. Overall, the activity level and the diversity metrics show a positive correlation, as shown in Figs.~\ref{fig:Entropy}(a) and~\ref{fig:Entropy}(b): more active editors tend to engage with a broader range of subjects more evenly. Consistently, the number of communities edited by an editor is also positively correlated with the diversity measures. By comparison, editing bots, indicated by black edges, exhibit high entropy and a wide range of inverse Simpson index values. This suggests that these bots operate by traversing a wide range of communities rather than focusing on specific subjects.

To formalize this diversity-based distinction, we establish a minimal non-trivial diversity criterion, defined as the case in which an editor contributes equally to two communities without preference, i.e., $p^{\alpha}_{c_1}=p^{\alpha}_{c_2}=1/2$ [see Eq.~\eqref{eq:pc_alpha}]. This yields two classification baselines, $H^*=\ln 2$ and $1/\lambda^*=2$ (we drop the index $\alpha$ for the baselines from now on). Figure~\ref{fig:Entropy}(c) illustrates the relationship between entropy $H$ and the inverse Simpson index $1/\lambda$, along with the baseline values $H^*$ and $1/\lambda^*$. Their strong positive correlation arises naturally, since $H$ and $1/\lambda$ are associated with the moments of $\ln p$ and $p$, respectively. Nevertheless, Fig.~\ref{fig:Entropy}(c) provides a useful basis for classifying editor types based on diversity. A \emph{`specialist'} is defined as an editor who edits intensively within a limited number of subject communities ($H_\alpha \le H^*$ and $1/\lambda_\alpha\le 1/\lambda^*$), whereas a \emph{`generalist'} is an editor whose activity is more evenly distributed across a wide range of communities ($H_\alpha > H^*$ and $1/\lambda_\alpha > 1/\lambda^*$). The specialists occupy a small region of the $1/\lambda$-$H$ plane but are densely populated. In total, we identify $14\,157$ specialists, $31\,254$ generalists, and 515 bots. We treat bots as a separate type from human editors, as they operate using automated algorithms rather than human-driven interests. We exclude editors with $H_\alpha > H^*$ and $1/\lambda_\alpha \le 1/\lambda^*$, as this regime is not of primary interest in our analysis.

\subsection*{Combined analysis of temporal and structural editing patterns via Jaccard similarity}
\label{subsec:result_Jaccard}

To integrate the temporal patterns captured by transition networks with the structural organization represented by the hyperlink network, we quantify their relationship at the level of individual editors using the Jaccard similarity. As defined in the Methods section, we calculate the Jaccard similarity $J_\alpha$ between each editor's \emph{transition} network $G_t^\alpha$ [Fig.~\ref{fig:Schematic}(c)] and the corresponding \emph{hyperlink} subnetwork $G_h^\alpha$, based on the overlap of their link sets. A high $J_\alpha$ value indicates strong structural overlap between the two networks, suggesting that an editor's transitions are more closely aligned with the hyperlink structure, whereas a low value indicates weaker overlap with the structure. The Jaccard similarity thus quantifies the extent to which editors' transition links overlap with the hyperlink subnetwork.

\begin{figure}[t!]
\centering
\includegraphics[width=\linewidth]{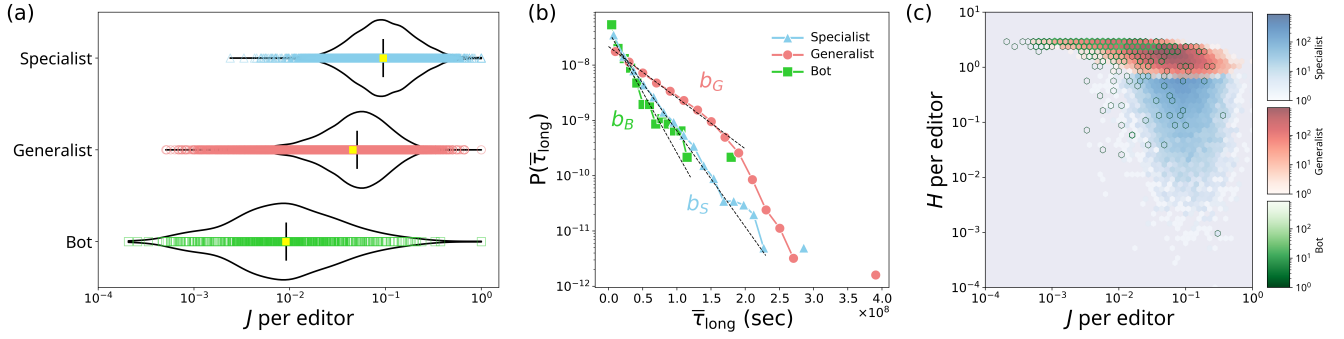}
\caption{Temporal-structural relationships of editorial behavior across editor types. 
(a) Violin plots of the Jaccard similarity $J$ by editor type. Each point represents an individual editor. The yellow box indicates the mean, and the vertical line represents the median.
(b) Distribution of the mean long-term IET per editor, $\bar{\tau}_{\rm long}$. The dashed lines indicate exponential tail fits of the form $\exp(-\bar{\tau}_{\rm long}/b)$, from which the characteristic scales are estimated as $b_S \sim 297$ days for `specialists', $b_G \sim 543$ days for `generalists', and $b_B \sim 228$ days for `bots'.
(c) Density plot of the Jaccard similarity $J$ versus entropy $H$ at the individual editor level. The blue and red color bars represent the densities of specialists and generalists, respectively. The density of bots is indicated by the color of the hexagon edges (green scale).
}
\label{fig:Jaccard} 
\end{figure}

In Fig.~\ref{fig:Jaccard}(a), the distributions of Jaccard indices differ across editor types. Specialists exhibit a narrower range than generalists and bots, while the latter two show broader and partially overlapping ranges. The distributions shift toward lower values from specialists to generalists to bots, as indicated by their mean and median values. These patterns suggest that specialists' transition networks are more closely aligned with the hyperlink structure, whereas bots exhibit weaker and more heterogeneous patterns, in line with their automated behavior. To further examine how structural tendencies relate to temporal behavior, we compare the Jaccard similarity with the mean IET and the entropy. Since transitions between hyperlinked pairs exhibit shorter IETs, as shown in the earlier IET analysis, specialists are expected to have shorter IETs if their transitions are more strongly aligned with the hyperlink structure. To examine this, we consider the mean long-term IET $\bar{\tau}_{\alpha, \,\rm long}$, averaged over all edited articles $\beta$ by an editor $\alpha$. The distributions, along with fitting lines, are shown in Fig.~\ref{fig:Jaccard}(b). Among human editors, specialists exhibit shorter mean long-term IETs ($\sim 297$ days) than generalists ($\sim 543$ days). Bots show the shortest mean long-term IETs, which differ from the human-editor trend and likely result from their automated editing behavior.

The relationship between entropy and the Jaccard similarity is shown in Fig.~\ref{fig:Jaccard}(c). While specialists are confined to the low-entropy regime ($H \leq H^*$) by definition, they are concentrated within a relatively narrow range of higher Jaccard values. By contrast, generalists span a broader range of Jaccard indices within the high-entropy regime, indicating that diversity in editing does not necessarily translate into stronger overlap with the hyperlink structure. This shows that entropy and the Jaccard similarity capture distinct aspects of editorial behavior. Together with the IET comparison, these results indicate that the Jaccard similarity provides complementary information to temporal and diversity-based measures.

Bots, however, are difficult to distinguish from generalists based on diversity alone (see Fig.~\ref{fig:Entropy}), as they often exhibit similarly high entropy. When the Jaccard similarity is considered, they predominantly occupy the low-Jaccard regime, indicating that their transitions are governed more by predefined algorithms than by the hyperlink structure. Some bots can exhibit low entropy and appear statistically similar to specialists. For example, the bot `PseudoBot', which focuses on date-related articles, shows extremely low entropy ($H \approx 0.001$) and inverse Simpson index ($1/\lambda \approx 1$), along with a relatively high Jaccard similarity ($J\approx 0.329$) due to the dense interlinking among such articles. Despite this statistical similarity, its behavior is algorithm-driven rather than based on human editorial preferences.

These observations highlight the need to interpret the three measures jointly. Among human editors, lower Jaccard similarity is associated with longer mean IETs, whereas bots depart from this tendency by showing low Jaccard similarity but the shortest mean IETs. This indicates that their rapid transitions are driven by automated procedures rather than hyperlink-based navigation. Thus, diversity measures, mean IETs, and Jaccard similarity together provide a more stable characterization of specialists, generalists, and bots, as summarized in Table~\ref{tab:class}.

\begin{table}[ht]
\centering
\caption{Summary of editor-type characteristics based on diversity, temporal, and structural measures.}
\label{tab:editor_type_summary}
\begin{tabular}{lccc}
\hline
\textbf{Editor Type} & \textbf{Topic Diversity} & \textbf{Mean IET} & \textbf{Jaccard Similarity} \\
\hline
Specialists 
& Low $H$; low $1/\lambda$ 
& Shorter than generalists 
& Relatively high; narrow range \\
Generalists 
& High $H$; high $1/\lambda$ 
& Longer than specialists 
& Lower and broader than specialists \\
Bots 
& Often high $H$; variable $1/\lambda$
& Shortest 
& Mostly low; heterogeneous \\
\hline
\end{tabular}
\label{tab:class}
\end{table}

\section*{Discussion}
\label{sdc:Discussion}

Wikipedia provides hyperlink connections between articles, a structure that supports navigation among semantically related topics. In this study, we examined whether this hyperlink structure is also associated with the editing trajectories of Wikipedia editors. We first found that hyperlinked article pairs exhibit higher transition density and shorter inter-event times (IETs) than sampled non-hyperlinked pairs, indicating that connected articles are effectively closer in editing sequences. This result suggests that hyperlinks may help identify related article pairs that are more readily connected in editors' subsequent editing activity. We then showed that this tendency differs across editor types. Specialists exhibit stronger similarity between their transition networks and the corresponding hyperlink subnetworks, whereas generalists show weaker and broader similarity patterns. Bots form a distinct class. Although they often exhibit high topical diversity, their low Jaccard similarity and shortest mean IETs indicate that their transitions are mainly governed by automated procedures rather than hyperlink-based navigation. These results show that topical diversity, temporal transition scale, and hyperlink-based structural similarity provide complementary information for characterizing editorial behavior.

These findings connect the temporal dynamics of editing with the structural organization of Wikipedia articles. Previous studies have often treated hyperlinks as navigational tools for readers or analyzed editor behavior through editor-article activity patterns. By combining edit histories with the article-to-article hyperlink network, our analysis shows that the existing knowledge structure is associated with how editors move between articles. In particular, the comparison between transition networks and hyperlink subnetworks reveals that hyperlink structure is not merely a static background of articles but is related to the sequential organization of editing activity. This perspective extends the conventional bipartite view of Wikipedia editing by explicitly incorporating article-to-article relations into the analysis of editor behavior. It also suggests that Wikipedia can be understood as an interdependent system in which article structure and editorial activity are coupled: hyperlinks organize knowledge at the article level, while editors move through and modify this structure through their editing trajectories.

There are several limitations to this study. First, our analysis is based only on the English Wikipedia, and similar analyses across different language editions would be needed to test the generality of the findings. Second, we used a static hyperlink snapshot together with dynamic edit histories. Because hyperlinks themselves evolve, future work should examine the co-evolution of editing trajectories and hyperlink structure. Third, the present study provides observational evidence for the association between hyperlink structure and editing behavior, but it does not establish a causal mechanism. Extending this framework with temporal hyperlink data or generative models of article selection would enable a more detailed understanding of how Wikipedia’s knowledge structure and editorial dynamics shape each other.
\section*{Methods}
\label{sec:method}

\subsection*{Wikipedia dataset}
\label{subsec:Dataset}

We utilized datasets from Wikimedia Downloads~\cite{wikimedia_dump}, specifically the English Wikipedia dump dated January 23, 2025, and selected five files: \texttt{stub-meta-history.xml}, \texttt{pagelinks.sql}, \texttt{page.sql}, \texttt{redirect.sql}, and \texttt{linktarget.sql}. First, we extracted the `edit-history dataset' from \texttt{stub-meta-history.xml}, which consists of page IDs, edit timestamps, and user IDs, while excluding redirect pages. Subsequently, we combined the \texttt{pagelinks.sql}, \texttt{page.sql}, \texttt{redirect.sql}, and \texttt{linktarget.sql} files to construct the `hyperlink dataset'. We used \texttt{linktarget.sql} to convert hyperlink information in \texttt{pagelinks.sql} and redirection information in \texttt{redirect.sql} from link target IDs to original page IDs. Additionally, \texttt{page.sql} was used to map these page IDs to their corresponding page titles. To ensure that all hyperlinks pointed to final destination pages rather than intermediate redirection pages, we updated any hyperlink targeting a redirection page to its final destination. Finally, we compiled a list of bot usernames from ``Category:All Wikipedia bots''~\cite{wiki_bots} to create an `automated bot list'. In the first step of preprocessing, we extracted $6\,943\,743$ pages and $780\,982\,746$ edits from the raw dataset, focusing on pages in the main \texttt{namespace} (i.e., \emph{articles} used in this study) and excluding all redirect pages. For anonymous users without a user ID, their IP addresses were used as user identifiers.

To observe representative editing behaviors of \emph{active editors}, we introduced three filtering criteria. First, we retained only articles present in the `hyperlink dataset', resulting in $6\,942\,548$ articles and $780\,961\,830$ edits. Next, we focused on the tail of the activity distribution (see Supplementary Fig.~S3) by restricting our analysis to articles and editors with more than $1\,000$ edits. Based on the dataset obtained from the first step, from the article perspective, this threshold retained approximately 1.5\% of the articles and 30.61\% of the edits, while from the editor perspective, it retained 0.08\% of the editors and 61.14\% of the edits. The intersection of these criteria yielded a final dataset of $104\,134$ articles, $51\,931$ editors, $14\,861\,487$ hyperlinks, and $117\,524\,906$ edits. Furthermore, we observed qualitatively similar tendencies when increasing the threshold to $2\,000$ or $3\,000$ edits (see Supplementary Figs.~S4,~S5 and Table~S1). Finally, Fig.~\ref{fig:Schematic}(a) provides a schematic overview of the dataset, including the relationships between editors and articles and the article-to-article hyperlink network. Specifically, editors contribute to multiple articles, and each article is edited by multiple users. Additionally, articles are interconnected via unidirectional or bidirectional hyperlinks.

\subsection*{Measurement of IETs in article transitions}
\label{sec:IET}
The IET is generally defined as the time interval between the occurrence of a preceding event and a subsequent event, a concept widely used to characterize the temporal dynamics of various human activities~\cite{barabasi2005origin, karsai2018bursty, choi2021individual}.
In this study, to analyze how the structure affects the behavior of editors, we specifically focus on the IETs measured during transitions between different articles.
To formalize this, let each edit be represented by a tuple $(\alpha,\beta,t)$, meaning that the editor $\alpha$ edited the article $\beta$ at time $t$. For a given editor, say $\alpha$, we derive the $\alpha$'s edit sequence in a chronological order, denoted by $\mathcal{B}_\alpha \equiv \{(\beta_1,t_1),\ldots,(\beta_n,t_n)\}$, where $t_i<t_j$ if $i<j$ [see Fig.~\ref{fig:Schematic}(b)]. Here $n$ is the number of edits by the editor $\alpha$. We first obtain the IET distribution by aggregating the time intervals between two \emph{consecutive} edits, i.e., $\tau_i\equiv t_{i+1}-t_i$ for $i=1,\ldots,n-1$, by ignoring the information on edited articles. We refer to this as the overall IET, denoted by $\tau_{\rm cons}$.

Next, let us denote the set of distinct articles edited by $\alpha$ by $V_\alpha$. Some articles in $V_\alpha$ contain hyperlinks referring or leading to other articles in $V_\alpha$. Such a \emph{hyperlink} structure can be represented by a directed, binary network $G_h^{\alpha}$, where we ignore self-loops. The set of nodes in $G_h^{\alpha}$ is $V_{\alpha}$ by definition, and the set of directed links in $G_h^{\alpha}$ is denoted by $E_{\alpha}$; if an article $\beta$ contains a hyperlink to another article $\beta'$ ($\neq \beta$), this referral makes a directed link $\beta\to\beta'$. Note that this hyperlink subnetwork plays a substrate of the editor's editing activity. Now the edit sequence $\mathcal{B}_\alpha$ can be seen as a hopping process on the network $G_h^{\alpha}$, while the hopping does not necessarily have to occur following the directed links in $G_h^{\alpha}$. Whenever $\beta_i\neq\beta_{i+1}$ happens, we call it an \emph{immediate} transition and then measure the \emph{immediate} IET defined by 
\begin{align}
    \tau_{\rm imme}\equiv \tau_{\beta_i\to\beta_{i+1}}=~t_{i+1}-t_i~.
\label{eq:tau_imme}
\end{align}
Then we obtain the set of such IETs only for the case when $~{\beta_i\to\beta_{i+1}}\in E_{\alpha}$, i.e.,
\begin{align}
    \mathcal{T}_{\rm imme,link}\equiv \{\tau_{\beta_i\to\beta_{i+1}} |~ \beta_i\to\beta_{i+1}\in E_{\alpha}\}.
    \label{eq:T_imme,link}
\end{align}
Similarly, we obtain the set of such IETs for all other cases, i.e., 
\begin{align}
    \mathcal{T}_{\rm imme,nonlink}\equiv \{\tau_{\beta_i\to\beta_{i+1}} |~ \beta_i\to\beta_{i+1}\notin E_{\alpha}\}.
    \label{eq:T_imme,nonlink}
\end{align}
Note that the hyperlinked case in Eq.~\eqref{eq:T_imme,link} includes the reciprocal hyperlinks; we obtain the subset of $\mathcal{T}_{\rm imme,link}$ only for the \emph{reciprocal} hyperlinks as follows:
\begin{align}
    \mathcal{T}_{\rm imme,reci-link}\equiv \{\tau_{\beta_i\to\beta_{i+1}} |~ \beta_i\to\beta_{i+1}\in E_{\alpha}\ \& \ \beta_{i+1}\to\beta_{i}\in E_{\alpha}\}.
    \label{eq:T_imme,reci-link}
\end{align}
For the uni-directionally hyperlinked articles, i.e., when $\beta_i\to\beta_{i+1}\in E_{\alpha}\ \& \ \beta_{i+1}\to\beta_{i}\notin E_{\alpha}$, one can separate the forward and backward transitions:
\begin{align}
    \mathcal{T}_{\rm imme,uni-link,forward} &\equiv \{\tau_{\beta_i\to\beta_{i+1}} |~ \beta_i\to\beta_{i+1}\in E_{\alpha}\ \& \ \beta_{i+1}\to\beta_{i}\notin E_{\alpha}\},\\
    \mathcal{T}_{\rm imme,uni-link,backward} &\equiv \{-\tau_{\beta_i\to\beta_{i+1}} |~ \beta_i\to\beta_{i+1}\notin E_{\alpha}\ \& \ \beta_{i+1}\to\beta_{i}\in E_{\alpha}\}.
    \label{eq:T_imme,uni-link}
\end{align}

To study such transition patterns on a longer timescale, we define the \emph{long-term} transition between a pair of articles. We choose two distinct articles in $V_{\alpha}$, say $\beta$ and $\beta'$. From the edit sequence $\mathcal{B}_\alpha$, we derive the edit sequence including only articles $\beta$ and $\beta'$ in a chronological order such that $\mathcal{B}_\alpha^{\beta,\beta'} \equiv \{(\beta_k,t_k)~|~\beta_k\in\{\beta,\beta'\}\}$, where $k=1,\ldots,m$. Whenever $\beta_k\neq\beta_{k+1}$, we measure the \emph{long-term} IET defined as
\begin{align}
    \tau_{\rm long}\equiv \tau_{\beta_k\to \beta_{k+1}}=~ t_{k+1}-t_k~.
\label{eq:tau_long}
\end{align}
Then we obtain the set of such IETs only for the case when $\beta_k\to\beta_{k+1}\in E_{\alpha}$, i.e.,
\begin{align}
    \mathcal{T}_{\rm long,link}\equiv \{\tau_{\beta_k\to\beta_{k+1}} |~ \beta_k\to\beta_{k+1}\in E_{\alpha}\}.
    \label{eq:T_long,link}
\end{align}
Similarly, we obtain the set of such IETs for all other cases, i.e., 
\begin{align}
    \mathcal{T}_{\rm long,nonlink}\equiv \{\tau_{\beta_k\to\beta_{k+1}} |~ \beta_k\to\beta_{k+1}\notin E_{\alpha}\}.
    \label{eq:T_long,nonlink}
\end{align}
Note that the hyperlinked case in Eq.~\eqref{eq:T_long,link} includes the reciprocal hyperlinks; we obtain the subset of $\mathcal{T}_{\rm long,link}$ only for the reciprocal hyperlinks as follows:
\begin{align}
    \mathcal{T}_{\rm long,reci-link}\equiv \{\tau_{\beta_k\to\beta_{k+1}} |~ \beta_k\to\beta_{k+1}\in E_{\alpha}\ \& \ \beta_{k+1}\to\beta_{k}\in E_{\alpha}\}.
    \label{eq:T_long,reci-link}
\end{align}
For the uni-directionally hyperlinked articles, i.e., when $\beta_k\to\beta_{k+1}\in E_{\alpha}\ \& \ \beta_{k+1}\to\beta_{k}\notin E_{\alpha}$, one can separate the forward and backward transitions:
\begin{align}
    \mathcal{T}_{\rm long,uni-link,forward} &\equiv \{\tau_{\beta_k\to\beta_{k+1}} |~ \beta_k\to\beta_{k+1}\in E_{\alpha}\ \& \ \beta_{k+1}\to\beta_{k}\notin E_{\alpha}\},\label{eq:T_long,uni-link,forward}\\
    \mathcal{T}_{\rm long,uni-link,backward} &\equiv \{-\tau_{\beta_k\to\beta_{k+1}} |~ \beta_k\to\beta_{k+1}\notin E_{\alpha}\ \& \ \beta_{k+1}\to\beta_{k}\in E_{\alpha}\}.
    \label{eq:T_long,uni-link,backward}
\end{align}

For both \emph{immediate} and \emph{long-term} transitions, the unique pairs are defined as the sets of distinct article combinations $\{\beta, \beta'\}$ actually utilized by an editor within each respective process. In this definition, a pair is treated as a single unique entity regardless of the transition direction or the specific orientation of the hyperlink. In \emph{immediate} transitions, a unique pair $\{\beta, \beta'\}$ is included only if it appears as a strictly consecutive edit $(\beta_i, \beta_{i+1})$ in the sequence $\mathcal{B}_\alpha$, where $\beta_i\neq\beta_{i+1}$ and $\{\beta_i, \beta_{i+1}\} = \{\beta, \beta'\}$. For \emph{long-term} transitions, the unique pairs consist of all pairs $\{\beta, \beta'\}$ selected to construct the filtered chronological sequences $\mathcal{B}_\alpha^{\beta, \beta'}$. These unique pairs are further categorized into hyperlinked and non-hyperlinked cases based on their existence in $E_\alpha$. Consequently, the number of unique pairs in immediate transitions is inherently less than or equal to that in long-term transitions, as the former is constrained by strict temporal adjacency in $\mathcal{B}_\alpha$.

\subsection*{Entropy and inverse Simpson index}
\label{sec:Entropy_inverseSimpsonindex}
To analyze and classify editor behavior, we utilize the concepts of entropy $H_\alpha$ and the inverse Simpson index ${1}/{\lambda_\alpha}$~\cite{simpson1949measurement}. These metrics quantify the degree to which an editor's edits are concentrated in or widely distributed across the communities within the hyperlinked network.

Let $e_c^{\alpha}$ be the number of edits by editor $\alpha$ within community $c$, and $e_c$ be the total number of edits in community $c$. The relative contribution of editor $\alpha$ to community $c$, denoted by $p_c^{\alpha}$, is defined as:
\begin{equation}
p_c^\alpha = \frac{(e_c^\alpha/e_c)}{\sum_{g \in C} (e_g^\alpha/e_g)}~,
\label{eq:pc_alpha}
\end{equation}
where $C$ is the set of all communities. 
Using $p_c^{\alpha}$, the entropy $H_\alpha$ is a metric utilized to quantify the diversity of an editor's community visit and edit activity, defined by the following equation:
\begin{equation}
H_\alpha = - \sum_{c \in C} p_c^\alpha \ln p_c^\alpha~.
\label{eq:entropy}
\end{equation}
A large $H_\alpha$ indicates that the editor has visited and edited across diverse communities in a balanced manner. Conversely, a small $H_\alpha$ implies that the editor has focused their visits and edits intensively on a limited number of communities.

To establish a stricter criterion for editor classification, the inverse Simpson index ${1}/{\lambda_\alpha}$ is introduced. This index signifies the effective number of communities to which the editor has made a substantial contribution, and it is defined as follows:
\begin{equation}
\frac{1}{\lambda_\alpha} = \frac{1}{\sum_{c \in C} (p_c^\alpha)^2}~.
\label{eq:inverse_Simpson_index}
\end{equation}

\subsection*{Jaccard similarity}
\label{sec:Jaccard}
To quantify the overlap between hyperlink connections and editors' article-to-article transitions, we measure the similarity between the \emph{transition} network $G_t^\alpha$ and the \emph{hyperlink} subnetwork $G_h^\alpha$ using the Jaccard similarity. For the construction of editor $\alpha$'s transition network $G_t^\alpha$, as schematically illustrated in Fig.~\ref{fig:Schematic}(c), we first identify the editor's effective edit activities.
For a given editor $\alpha$, we first construct a \emph{transition} network $G_t^\alpha=(V_\alpha, E_t^\alpha)$, where $V_\alpha$ is the set of articles edited by $\alpha$. To identify effective editing transitions, we apply the burst-train technique~\cite{karsai2012universal}. A single burst is a sequence of consecutive timestamps where the IET between adjacent events does not exceed the threshold $\Delta t$. If the IET exceeds the threshold $\Delta t$, the influence of the editing trajectory is considered terminated. In this study, we set $\Delta t$ to 1 day; the robustness and sensitivity of this threshold across different editor types are detailed in Supplementary Note 4.
The set of transition links $E_t^\alpha$ is constructed by identifying consecutive article transitions within the same burst:
\begin{equation}
    E_t^\alpha=\{(\beta_i, \beta_{i+1}) \mid~ t_{i+1}-t_i \le \Delta t,~ \beta_i\neq\beta_{i+1}\},
\label{eq:transition_links}
\end{equation}
where $\beta_i$ and $\beta_{i+1}$ are articles in the edit sequence $\mathcal{B}_\alpha$. To simplify the comparison, we treat $E_t^\alpha$ as a set of binary, undirected links, disregarding self-loops, directionality, and weights. We confirmed that while considering directionality results in lower Jaccard similarity values, the overall relative tendencies across editor types remain consistent with the undirected case.

Next, we consider the corresponding \emph{hyperlink} subnetwork $G_h^\alpha=(V_\alpha, E_\alpha)$ on the same set of nodes. Although the link set $E_\alpha$ is directed, we disregard its directionality in this Jaccard similarity analysis to ensure consistency with the undirected transition link set $E_t^\alpha$.
Finally, the Jaccard similarity $J_\alpha$ for editor $\alpha$ is measured using the link sets $E_t^\alpha$ and $E_\alpha$ as follows:
\begin{equation}
    J_\alpha = \frac{|E_t^\alpha \cap E_\alpha|}{|E_t^\alpha \cup E_\alpha|}~.
\label{eq:Jaccard}
\end{equation}
A high $J_\alpha$ indicates that the editor's actual editing trajectory exhibits high similarity with the existing hyperlink connection structure. Conversely, a low $J_\alpha$ signifies that the editor's trajectory has low similarity to the hyperlink structure, suggesting a higher tendency to move to non-hyperlinked articles.

\subsection*{Use of an artificial intelligence tool}
In this study, a generative artificial intelligence (AI) tool, Google Gemini, was used solely to extract representative keywords from the titles of articles constituting each detected community in the Wikipedia hyperlink network. Specifically, for each of the 19 communities identified by the Leiden algorithm, the list of article titles belonging to that community was provided as input to the model, and the model was prompted to generate a concise set of representative topical keywords characterizing each community. The AI-generated keywords were subsequently reviewed and verified by the authors to ensure that they accurately reflected the thematic content of the corresponding communities. No generative AI tool was used in the writing or interpretation of the manuscript beyond this specific analytical task. The authors take full responsibility for the accuracy and integrity of all results reported in this study.

\section*{Data availability}

The raw data used in this study were obtained from the English Wikipedia dump (January 23, 2025) and are available at the Wikimedia Downloads repository (\url{https://dumps.wikimedia.org/}). These data sources are also cited in the References~\cite{wikimedia_dump}.


\section*{Acknowledgements}
We thank Dr.~Jong-Min Park for the fruitful discussion. This work was supported by the National Research Foundation (NRF) of Korea through Grant Numbers. RS-2024-00341317 (M.J.L.), RS-2026-25488703 (S.-W.S.), and RS-2026-25476645 (H.-H.J.).
Y.M. acknowledges support from JSPS KAKENHI Grant Number JP25K07145 and from RIKEN Pioneering Project ``Planetary Resilience Science for Safeguarding the Global Commons.''
We thank APCTP, Pohang, Korea, for their hospitality during the Topical Research Program [APCTP-2025-T04], from which this work greatly benefited. We also acknowledge the hospitality at APCTP, where part of this work was done. This work is partially supported by RIKEN R-CCS International Student Internship.

\section*{Author contributions statement}
Y.S., M.J.L., H.-H.J., S.-W.S., and Y.M. conceived the study and designed the research methodology. Y.S. collected the datasets and performed the numerical experiments and data analysis. Y.S., M.J.L., H.-H.J., S.-W.S., and Y.M. interpreted the results. Y.S. wrote the original draft of the manuscript. M.J.L., H.-H.J., S.-W.S., and Y.M. supervised the research and reviewed the manuscript. All authors reviewed and approved the final version of the manuscript.

\section*{Additional information}

\textbf{Supplementary Information} The online version contains supplementary material available at [Journal URL will be inserted by the publisher].

\textbf{Competing interests} The authors declare no competing interests.

\textbf{Correspondence} and requests for materials should be addressed to M.J.L., S.-W.S., H.-H.J., or Y.M.

\textbf{Code availability} \url{https://github.com/YeonjiSeo/wiki-hyperlink-editing}.

\clearpage
\setcounter{figure}{0}
\renewcommand{\thefigure}{S\arabic{figure}}
\setcounter{table}{0}
\renewcommand{\thetable}{S\arabic{table}}

\section*{Supplementary Information}

\subsection*{Supplementary Note 1: Structural properties of the hyperlink network}

\begin{figure}[b!]
   \centering
   \includegraphics[width=0.9\linewidth]{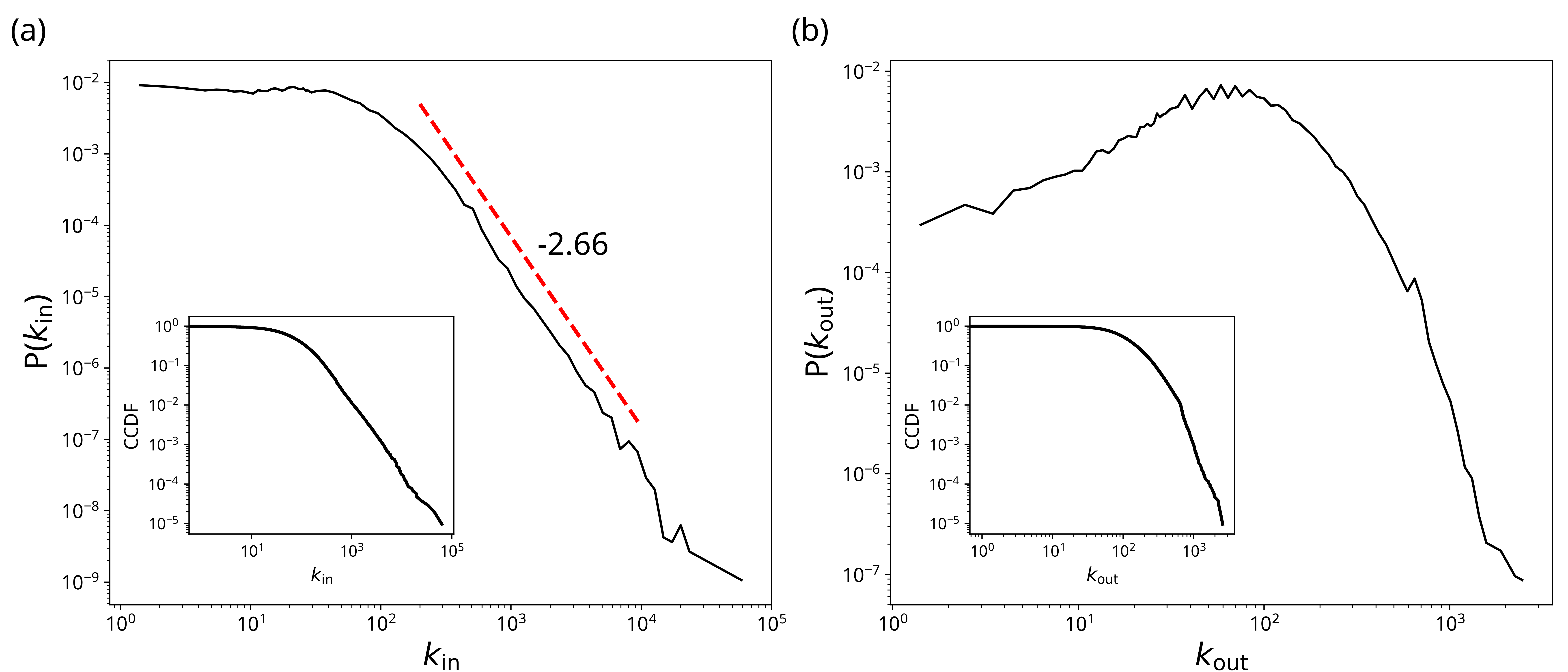}
   \caption{Degree distributions of the hyperlink network. (a) Distribution of the in-degree ($k_{\text{in}}$) and (b) out-degree ($k_{\text{out}}$) of the Wikipedia hyperlink network $G_h$. In both panels, the black solid lines denote the log-binned distributions. In panel (a), the red dashed line represents the power-law fit $P(k_{in}) \sim k_{in}^{-\gamma}$ with $\gamma = 2.66$. Insets: Complementary cumulative distribution functions (CCDFs).
}
   \label{fig:SI_degree}
\end{figure}

To understand the macroscopic structure of the Wikipedia knowledge map, we examined the degree distributions of the hyperlink network, $G_h$. Supplementary Fig.~\ref{fig:SI_degree} presents the probability distributions for in-degree, $k_{\text{in}}$, and out-degree, $k_{\text{out}}$. The in-degree represents the number of incoming hyperlinks to an article, indicating how frequently a specific topic is referenced by other articles for supplementary information. As shown in Supplementary Fig.~\ref{fig:SI_degree}(a), the majority of articles possess an in-degree of less than 100, where the distribution remains relatively flat before a sharp decline. However, a small number of articles exhibit an in-degree exceeding $1\,000$, functioning as central information hubs within the network.

The out-degree denotes the number of outgoing hyperlinks from an article, reflecting the density of references provided to explain concepts within that article. The distribution in Supplementary Fig.~\ref{fig:SI_degree}(b) shows a prominent peak around $k_{\text{out}} \approx 100$. The lower frequency of articles with very small out-degrees suggests that a typical Wikipedia article is structured to include approximately 100 internal hyperlinks as a standard for navigational and contextual depth. Similar to the in-degree distribution, a heavy tail exists where certain comprehensive articles contain more than $1\,000$ outgoing links. 

\subsection*{Supplementary Note 2: Relationship between editorial activity and diversity metrics.}

\begin{figure}[ht]
   \centering
   \includegraphics[width=\linewidth]{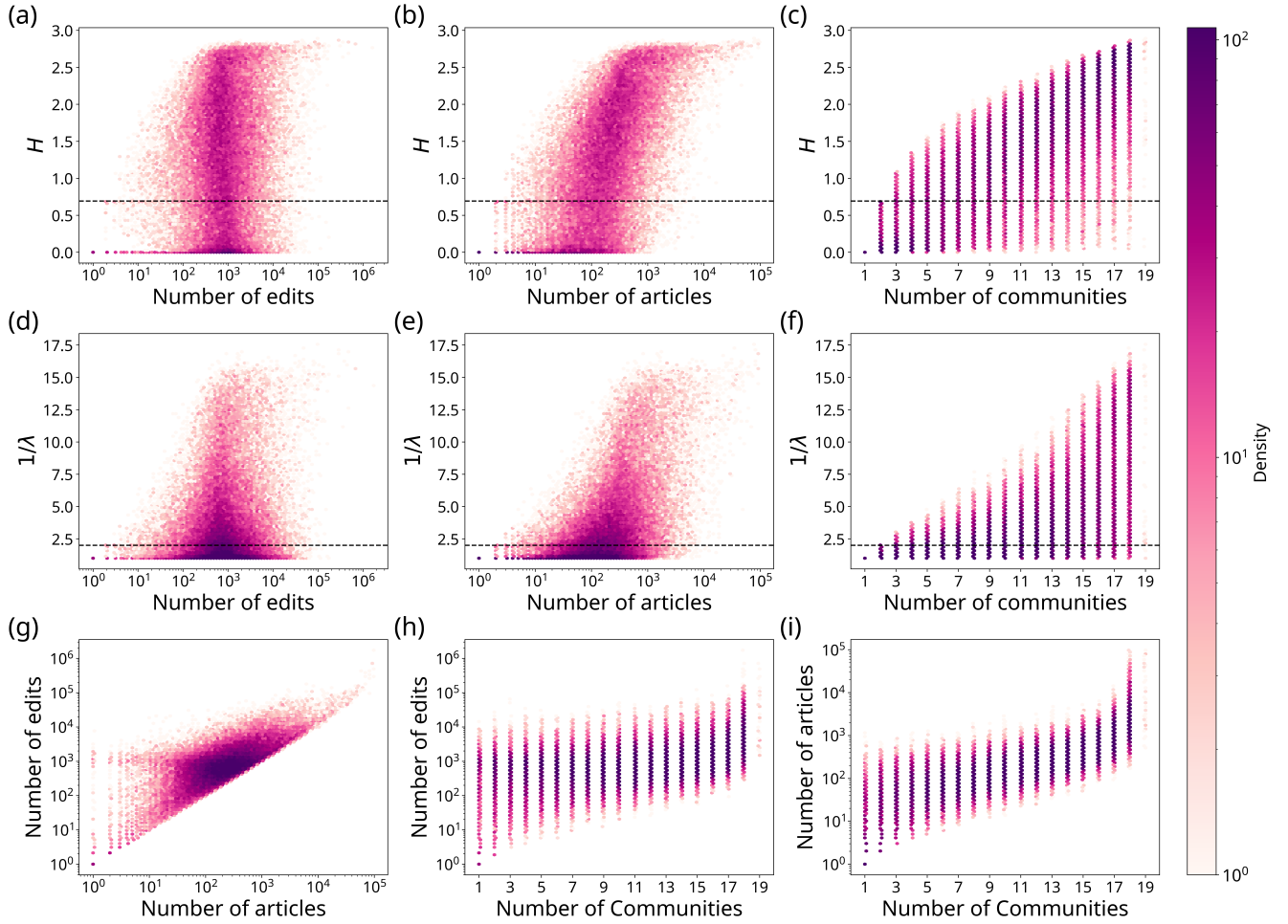}
   \caption{Density plots of editorial activity and diversity metrics measured per editor. The colorbar indicates the number of editors in each bin. (a) to (c) show the relationship between editorial entropy $H$ and the number of edits, articles, and communities, respectively. The horizontal dashed lines in these panels denote the classification threshold of $H^*=\ln{2}$. (d) to (f) display the relationship between the inverse Simpson index ${1}/{\lambda}$ and the same three activity metrics with horizontal dashed lines marking the threshold of ${1}/{\lambda^*}=2$. (g) to (i) present the pairwise correlations between editorial activity metrics, specifically comparing articles and edits in (g), communities and edits in (h), and communities and articles in (i).
}
   \label{fig:SI_values}
\end{figure}

We examined the relationship between an editor's activity level and their interest diversity by comparing the number of edits, articles, and communities with entropy and the inverse Simpson index.
As shown in Supplementary Figs.~\ref{fig:SI_values}(a)~and~\ref{fig:SI_values}(b), the majority of editors are concentrated around $1\,000$ edits, yet their entropy values are distributed across a wide range. For the number of articles, most editors fall within the 100 to $1\,000$ range and similarly exhibit diverse entropy levels. A comparison between entropy and the number of visited communities reveals that entropy tends to increase as editors engage with more communities, as illustrated in Supplementary Fig.~\ref{fig:SI_values}(c).
While the inverse Simpson index generally follows a trend similar to entropy, its distribution shows a distinct characteristic. Unlike entropy, which is relatively evenly spread, the inverse Simpson index is more densely concentrated at lower values as seen in Supplementary Figs.~\ref{fig:SI_values}(d),~\ref{fig:SI_values}(e),~and~\ref{fig:SI_values}(f). This suggests that even when editors visit multiple communities, their actual contributions are often focused on a few specific areas.
Comparing the number of edits and articles with the number of visited communities, we confirm that an increase in the number of edits and the number of articles edited generally leads to an expansion in the number of communities an editor participates in. These pairwise correlations are further detailed in Supplementary Figs.~\ref{fig:SI_values}(h)~and~\ref{fig:SI_values}(i).

\subsection*{Supplementary Note 3: Data selection and robustness of filtering thresholds}

\begin{figure}[b!]
   \centering
   \includegraphics[width=\linewidth]{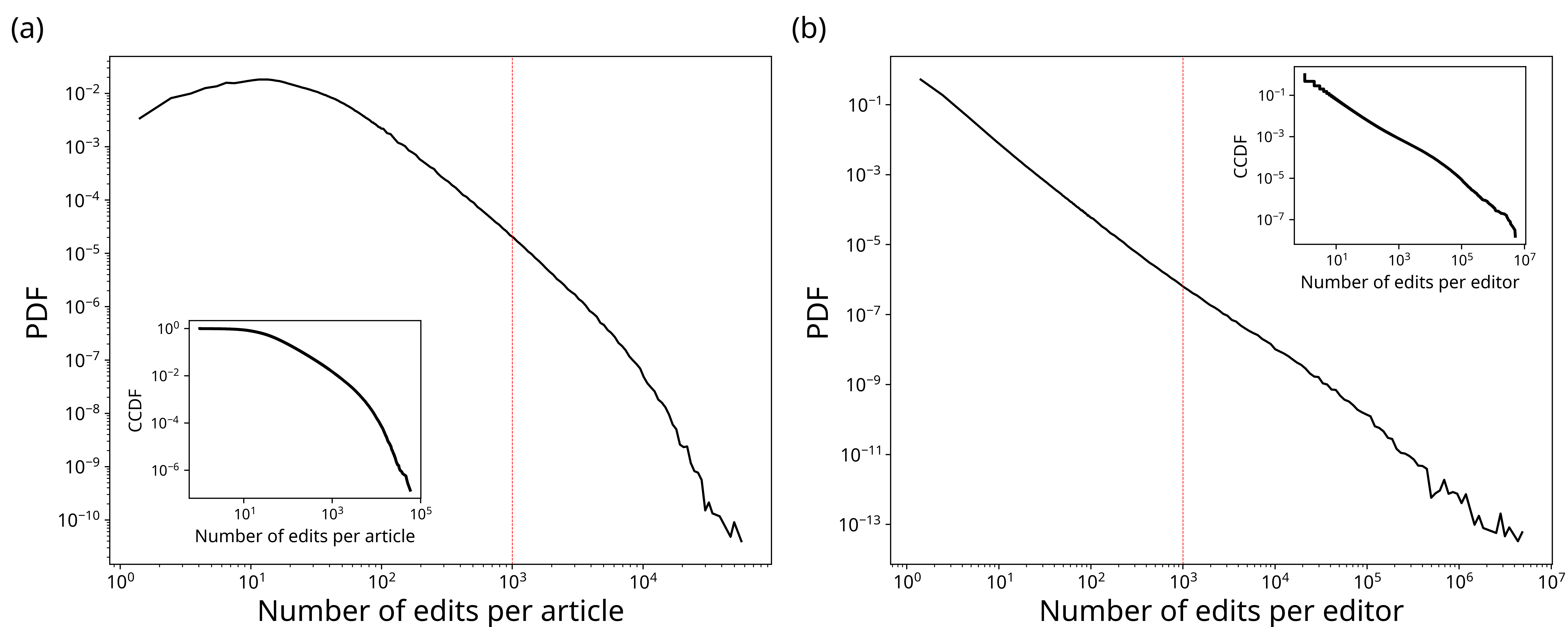}
   \caption{Statistical distributions of editorial activities. (a) Probability density function (PDF) of the number of edits per article. (b) PDF of the number of edits per editor. In both panels, the black solid lines denote the log-binned distributions. The insets show the corresponding complementary cumulative distribution functions (CCDF). The red dotted lines indicate the activity threshold of 1,000 edits.
}
   \label{fig:SI_distribution}
\end{figure}

To characterize the distribution of editorial activities, we examined the probability density functions (PDF) of edits for both articles and editors. As shown in Supplementary Fig.~\ref{fig:SI_distribution}, the number of edits per article (a) and per editor (b) exhibit a power-law-like decay over some range, followed by a long tail.
To exclude transient or one-time edits and focus on active participants within the Wikipedia ecosystem, we established an activity threshold of $1\,000$ edits. As illustrated by the red dotted lines in Supplementary Fig.~\ref{fig:SI_distribution}, this threshold effectively allows us to focus on the heavy-tail region where significant and sustained editorial actions occur. While only approximately 0.08\% of the total editors meet this criterion, they account for 61.14\% of the total editorial volume. 

\begin{table}[ht]
\caption{Dataset statistics according to different activity thresholds. The table summarizes the total number of edits, articles, editors, and hyperlinks retained as the activity threshold increases from $1\,000$ to $3\,000$ edits. The counts for each editor category, including specialist, generalist, and bot, are also provided for each threshold.
}
\centering
\begin{tabular}{|c|c|c|c|c|c|c|c|}
\hline
Threshold & Edits & Articles & Editors & Hyperlinks & Specialist & Generalist & Bot\\
\hline
1000 & $117\,524\,906$ & $104\,134$ & $51\,931$ & $14\,861\,487$ & $14\,157$ & $31\,254$ & $515$\\
\hline
2000 & $64\,984\,820$ & $38\,082$ & $30\,029$ & $4\,830\,227$ & $7\,785$ & $18\,449$ & $449$\\
\hline
3000 & $42\,274\,013$ & $19\,758$ & $21\,666$ & $2\,206\,548$ & $5\,980$ & $12\,816$ & $419$\\
\hline
\end{tabular}
\label{tab:SI_activity_thresholds}
\end{table}

In order to ensure that the $1\,000$ edit threshold does not introduce bias into our findings, we performed a robustness check by increasing the threshold to $2\,000$ and $3\,000$ edits. As summarized in Supplementary Table~\ref{tab:SI_activity_thresholds}, although the absolute numbers of articles and editors decrease as the threshold becomes more stringent, the fundamental characteristics of the dataset, such as the composition of specialist, generalist, and bot categories, remain consistent.

\begin{figure}[b!]
   \centering
   \includegraphics[width=0.9\linewidth]{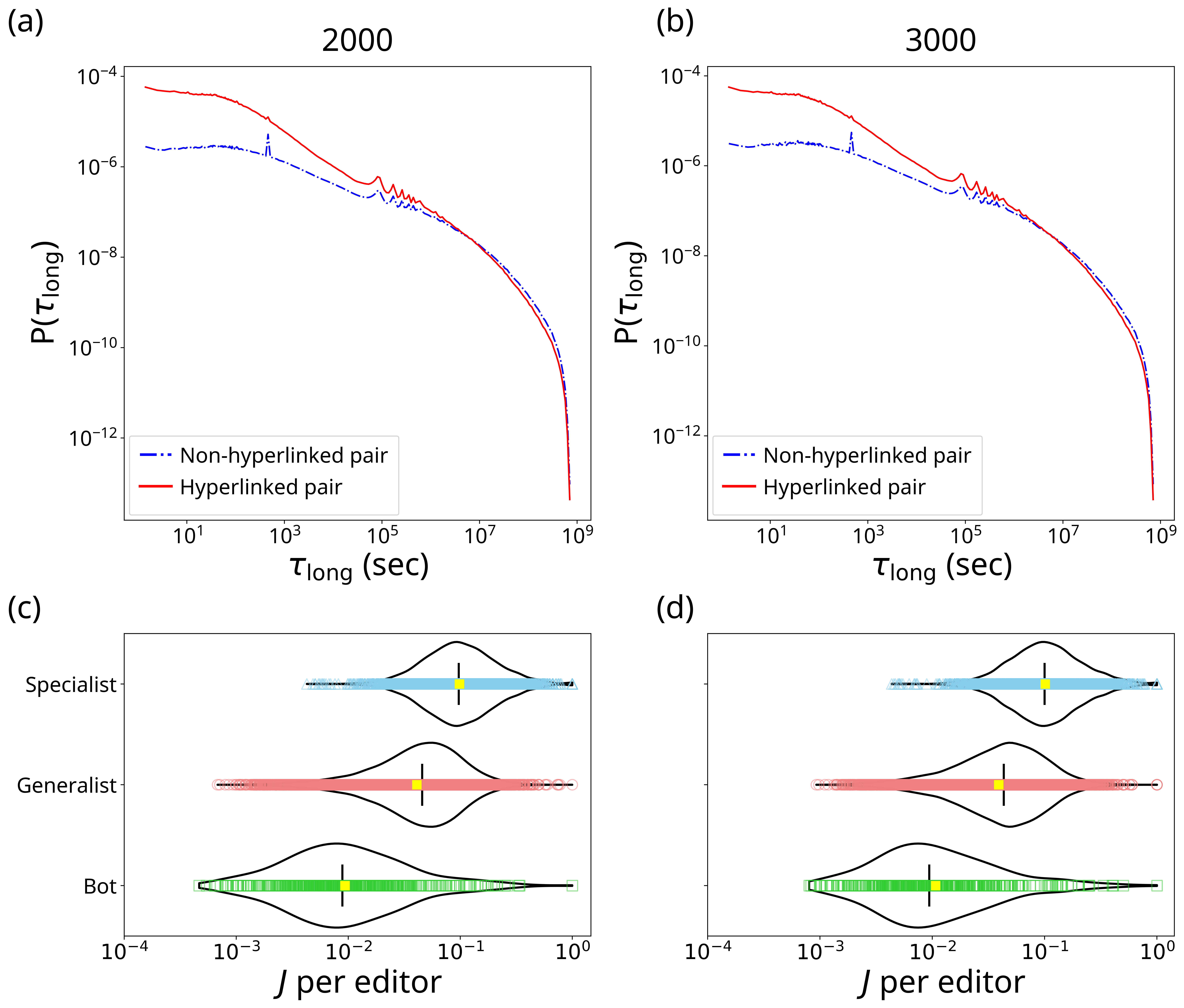}
   \caption{Robustness check of editorial patterns with higher activity thresholds. (a, b) Probability distribution of the long-term IET $\tau_{\rm long}$ for thresholds of (a) $2\,000$ and (b) $3\,000$ edits. Red solid lines represent hyperlinked pairs, while blue dash-dot lines represent non-hyperlinked pairs. (c, d) Jaccard similarity distribution by editor type for thresholds of (c) $2\,000$ and (d) $3\,000$ edits. The yellow boxes indicate the average values, and the vertical lines represent the median.
}
   \label{fig:SI_robustness}
\end{figure}

This consistency is further confirmed by the analytical results presented in Supplementary Fig.~\ref{fig:SI_robustness}. Even with higher thresholds of $2\,000$ edits (a, c) and $3\,000$ edits (b, d), the observed patterns remain identical to the main results. Hyperlinked article pairs are edited faster than non-hyperlinked pairs (a, b), and the Jaccard similarity distributions across editor types (c, d) maintain the same trends. These findings demonstrate that our results are robust and not dependent on a specific threshold selection.

\subsection*{Supplementary Note 4: Sensitivity of Jaccard similarity to the time threshold $\Delta t$}

\begin{figure}[b!]
   \centering
   \includegraphics[width=\linewidth]{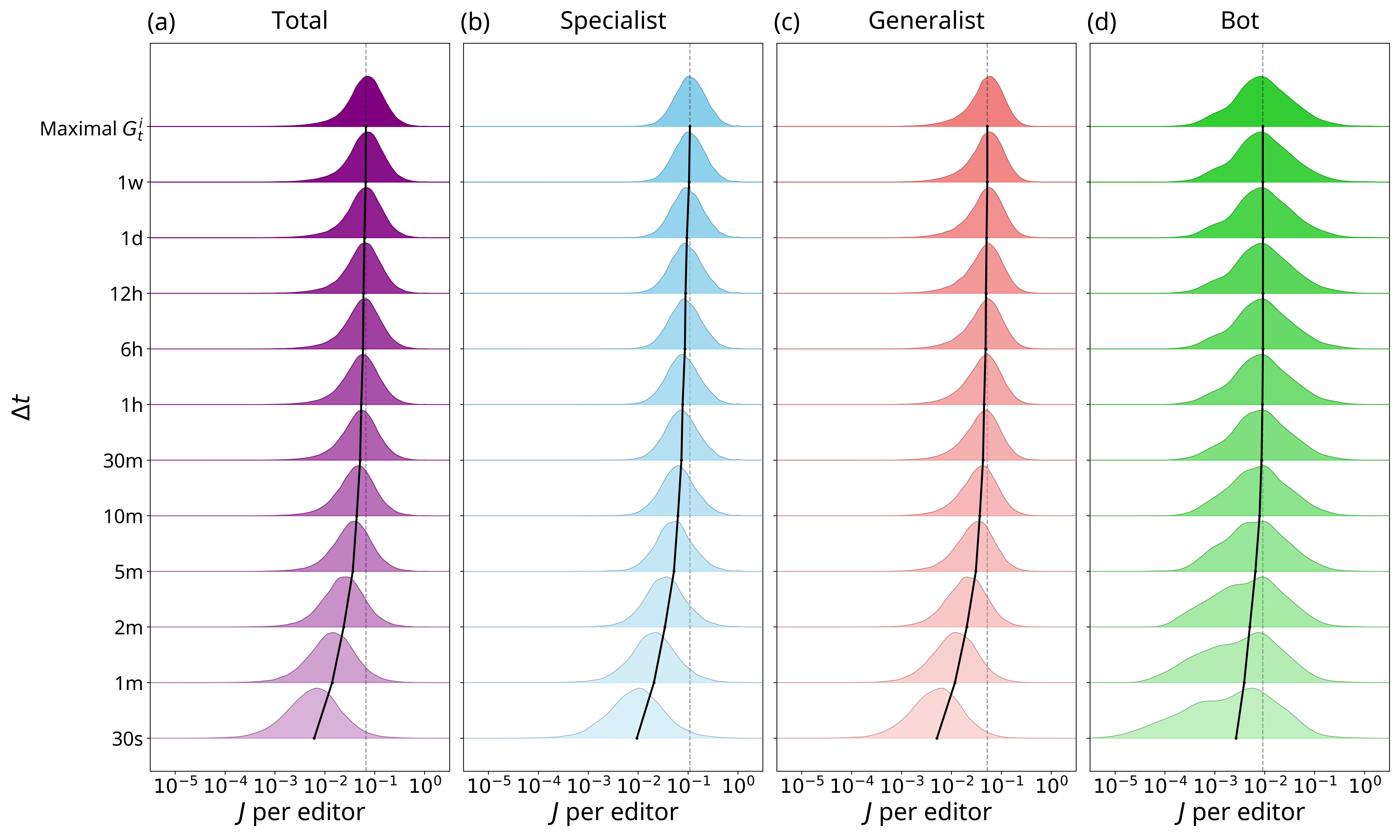}
   \caption{Sensitivity of Jaccard similarity distributions to the time threshold $\Delta t$. The ridge plots show the probability density distributions of the Jaccard similarity $J_\alpha$ for (a) Total, (b) specialist, (c) generalist, and (d) bot across varying $\Delta t$. The y-axis represents $\Delta t$ values from 30 seconds to the maximum inter-event time, Maximal $G_t^{\alpha}$. In each panel, the solid black line connects the median values, and the vertical dashed line indicates the median Jaccard similarity at the maximum $G_t^{\alpha}$.
}
   \label{fig:SI_sensitivity}
\end{figure}

When constructing the transition network from the edit trajectory, we defined and utilized $\Delta t$ as the maximal allowable inter-event time between two consecutive edit events. The setting of $\Delta t$ is a key parameter that determines whether an editor's transition behavior is regarded as a single consecutive flow or partitioned into distinct activity sessions. Therefore, we analyzed how changes in $\Delta t$ affect the distribution of the Jaccard similarity and the editorial characteristics across different editor types.

Supplementary Fig.~\ref{fig:SI_sensitivity} shows the Jaccard similarity distributions measured by varying $\Delta t$ from 30 seconds to the level of the maximal $G_t^{\alpha}$, where the entire edit history of an editor is connected into a single network. Across the entire range of $\Delta t$, the Jaccard similarity distributions for specialists, generalists, and bots maintain distinct patterns that are clearly separated from one another. Especially when comparing the results based on the vertical dashed lines, which represent the median values of the maximal $G_t^{\alpha}$, the tendency to follow the hyperlink structure is highest for specialists and lowest for bots, and this result is robustly maintained regardless of the $\Delta t$ setting. Looking at the solid black lines that connect the median values of each distribution, the Jaccard similarity values gradually increase as $\Delta t$ increases, showing a tendency to saturate after a certain point. For the human editor groups, the median Jaccard similarity is already close to that of the maximal $G_t^{\alpha}$ even when $\Delta t$ is set to 12 hours. In contrast, for bots, the distribution reaches a state similar to the maximal $G_t^{\alpha}$ with a much shorter $\Delta t$ of 10 minutes. Based on these analysis results, we set $\Delta t$ to 1 day for our main study.

\end{document}